\documentclass[prd,onecolumn,nofootinbib,amsmath,amssymb]{revtex4-2}

\usepackage[utf8]{inputenc}
\usepackage[T1]{fontenc}
\usepackage{lmodern}
\usepackage{bm}
\usepackage{hyperref}
\usepackage{graphicx}

\hypersetup{
    colorlinks=true,
    linkcolor=blue,
    citecolor=blue,
    urlcolor=blue,
}

\begin{document}

\title{A dynamical systems approach to studying the equivalence principle in dilaton gravity}

\author{Vel\'asquez-Toribio A.M.}
\email{alan.toribio@ufes.br}
\affiliation{Center for Astrophysics and Cosmology of the Esp\'{\i}rito Santo Federal University (UFES), Brazil.}

\date{\today}

\begin{abstract}
We study a string-inspired dilaton cosmology in the Damour--Polyakov (DP) regime using dynamical-systems methods,
aiming to make explicit how cosmological relaxation controls deviations from the equivalence principle.
Working in the Einstein frame, we consider a spatially flat FLRW universe filled with pressureless matter and a
universally coupled dilaton. Expanding the conformal coupling function and the scalar potential around the
least-coupling point, we obtain a closed and self-consistent autonomous system governing the late-time evolution of
the scalar-matter sector.
The resulting phase space contains a stable fixed point associated with least coupling, approached only
asymptotically along cosmological trajectories. Therefore, at any finite epoch the solution typically retains a
small displacement from the fixed point. In the DP regime this finite-epoch displacement sets the ambient coupling,
and thus determines the magnitude of fifth-force effects and deviations from the equivalence principle in the
nonrelativistic limit.
By linearising the system around a finite-epoch reference state, we show that the damping of the displacement is
controlled by the Jacobian eigenvalues of the DP fixed point. This yields a direct dynamical estimate of how rapidly
deviations from the equivalence principle are reduced during cosmological evolution. The mechanism is global and
cosmological in origin, and is conceptually distinct from local environmental screening as in chameleon or symmetron
scenarios. Overall, our results illustrate how phase-space techniques provide a clear bridge between cosmological
dynamics and weak-field departures from General Relativity.
\end{abstract}

\maketitle

\section{Introduction}
\label{sec:intro}

The dilaton field emerges naturally in string theory as the scalar partner of the graviton in the low energy effective action \cite{DamourPolyakov1994,DamourNordtvedt1993,DamourPiazzaVeneziano2002PRL,DamourPiazzaVeneziano2002PRD,FujiiMaeda}. After compactification to four dimensions and rewriting the theory in the Einstein frame, the dilaton appears as a real scalar field that couples to matter nonminimally through a conformal rescaling of the metric in the matter action. This structure modifies the gravitational dynamics, since the dilaton mediates an additional interaction whose strength depends on the value of the field and on the shape of its effective potential. As the dilaton relaxes toward the minimum of this potential, a point where its dynamics becomes energetically favored and its coupling to matter may diminish, the theory smoothly approaches a regime that can approximate General Relativity, while still allowing for departures from it when the field is displaced from equilibrium.

In this way, when the field is still approaching its equilibrium configuration but has not fully settled at the minimum, a variety of novel physical effects may appear. In particular, slight displacements from the equilibrium state can induce departures from the Equivalence Principle, opening the door to a broad and testable range of cosmological signatures \cite{BransDicke1961,Maeda1989,FujiiMaeda,Wetterich1988,GasperiniVeneziano1993,LidseyWandsCopeland2000,Tsujikawa2001,BeanMagueijo2001,GasperiniPiazzaVeneziano2002,PiazzaTsujikawa2004,BiswasEtAl2006,CatenaMoeller2008,Gasperini2007_PBBreview,Gasperini2008_DilatonCosmology,AkarsuDereli2013,Martins2015,Martins2019,Vacher2023,GrumillerKummerVassilevich2002}.
.

On the other hand, dynamical systems methods have long offered a powerful framework for analysing cosmological models and understanding their global behaviour. One of their main advantages is that they allow the full set of cosmological evolution equations to be rewritten in terms of dimensionless variables, making it possible to characterise entire families of solutions without specifying initial conditions in detail. This approach also provides a unified language for identifying fixed points, assessing their stability, and determining whether they correspond to physically relevant regimes such as matter domination, accelerated expansion, or scaling solutions. As a result, dynamical-systems techniques make it possible to distinguish generic behaviours from fine-tuned ones and to reveal attractor solutions that naturally emerge in wide classes of theories.
A vast body of work applies these tools to FLRW universes , to models exhibiting accelerated expansion , to anisotropic Bianchi geometries, and even to scenarios with chaotic dynamics, among many other contexts \cite{refe,WainwrightEllis1997,Coley2003,CopelandSamiTsujikawa2006,AmendolaTsujikawa2010,Boehmer2015,CopelandLiddleWands1998,BarreiroCopelandNunes2000,NunesMimoso2000,GuoPiaoZhangZhang2003,LeonSaridakis2013,Roy2014,TamaniniThesis2014,IvanovProdanov2019,Landim2019,Rathore2023,AlhoQuintessence2023}
.

Therefore, since our cosmological dilaton model admits an equilibrium point that plays a central role in its evolution, it is natural to adopt dynamical systems techniques. The phase-space structure thus provides a clear geometric picture that links the cosmological evolution of the dilaton to possible deviations from geodesic motion for test bodies.
In addition, we compare our results with two prominent screening scenarios, namely the symmetron and chameleon models, emphasizing both the similarities and the differences in how their effective potentials, coupling functions, and equilibrium configurations control the appearance and magnitude of deviations from the Equivalence Principle \cite{BraxMotaShaw2011,BraxBurrage2014_review,HinterbichlerKhoury2010_PRL,Hinterbichler2011_SymmetronCosmology,Davis2012_StructureSymmetron,Koivisto2012_ScreeningDisformal,Hagala2016_DisformalSymmetron,BraxSymmetronReview2015}.

The present paper is organized as follows: Section II presents the action and the corresponding field equations. Section III develops the autonomous dynamical system and examines their implications. Section IV investigates the implications for the equivalence principle. Section V presents some numerical results, and Section VI presents our conclusions and outlines possible directions for future work.

\section{The basic dilaton model}
\label{sec:dilaton-basic}

In this section we summarise a generic dilaton gravity model written in the Einstein frame, with a single real scalar field universally coupled to matter through a conformal factor. We use units $c=\hbar=1$ and metric signature $(-,+,+,+)$. The action is
\begin{equation}
S
=
\int d^4x\,\sqrt{-g}\,
\left[
\frac{M_{\rm Pl}^2}{2}\,R
-\frac{1}{2}\,g^{\mu\nu}\nabla_\mu\phi\,\nabla_\nu\phi
- V(\phi)
\right]
+
S_m\!\left[\psi_i,\,\tilde g_{\mu\nu}\right],
\label{eq:action}
\end{equation}
where $g_{\mu\nu}$ is the Einstein-frame metric, $M_{\rm Pl}$ is the reduced Planck mass, $\phi$ is the dilaton, and $V(\phi)$ is its potential. Matter fields $\psi_i$ couple minimally to the Jordan-frame metric
\begin{equation}
\tilde g_{\mu\nu}=A^2(\phi)\,g_{\mu\nu},
\label{eq:jordan-metric}
\end{equation}
with $A(\phi)>0$ a dimensionless conformal factor. By construction, the matter action $S_m$ depends on $\phi$ only through $\tilde g_{\mu\nu}$. In the Jordan frame, freely falling structureless test bodies follow geodesics of $\tilde g_{\mu\nu}$, so physical rods and clocks are naturally associated with $\tilde g_{\mu\nu}$.

The strength of the scalar coupling is controlled by
\begin{equation}
\alpha(\phi)\equiv \frac{d\ln A(\phi)}{d\phi}
=\frac{1}{A(\phi)}\frac{dA(\phi)}{d\phi}.
\label{eq:alpha-def}
\end{equation}
For example, $A(\phi)=e^{\beta\phi/M_{\rm Pl}}$ gives $\alpha(\phi)=\beta/M_{\rm Pl}$, whereas models with a minimum of $A(\phi)$ yield $\alpha(\phi_\star)=0$, providing a dynamical route to suppress scalar effects. The function $\alpha(\phi)$ will be central when discussing equivalence-principle tests: in the Einstein frame it controls the effective fifth force sourced by gradients of $\phi$, while in the Jordan frame the same physics is encoded in the metric $\tilde g_{\mu\nu}$.

Varying Eq.~\eqref{eq:action} with respect to $g^{\mu\nu}$ yields the Einstein equations
\begin{equation}
M_{\rm Pl}^2\,G_{\mu\nu}
=
T^{(\phi)}_{\mu\nu}+T^{(m)}_{\mu\nu},
\label{eq:einstein-eq}
\end{equation}
where the scalar energy-momentum tensor is
\begin{equation}
T^{(\phi)}_{\mu\nu}
=
\nabla_\mu\phi\,\nabla_\nu\phi
-
g_{\mu\nu}
\left[
\frac{1}{2}\,g^{\alpha\beta}\nabla_\alpha\phi\,\nabla_\beta\phi
+
V(\phi)
\right],
\label{eq:Tphi}
\end{equation}
and the matter energy-momentum tensor in the Einstein frame is defined by
\begin{equation}
T^{(m)}_{\mu\nu}
\equiv
-\frac{2}{\sqrt{-g}}\frac{\delta S_m}{\delta g^{\mu\nu}}\,.
\label{eq:Tm-def}
\end{equation}
It is sometimes useful to introduce also the Jordan-frame tensor
\begin{equation}
\tilde T^{(m)}_{\mu\nu}
\equiv
-\frac{2}{\sqrt{-\tilde g}}\frac{\delta S_m}{\delta \tilde g^{\mu\nu}}\,,
\end{equation}
which is covariantly conserved with respect to $\tilde\nabla_\mu$ because matter is minimally coupled to $\tilde g_{\mu\nu}$. The two tensors are related by
\begin{equation}
T^{(m)}_{\mu\nu}=A^2(\phi)\,\tilde T^{(m)}_{\mu\nu},
\qquad
T_{(m)}\equiv g^{\mu\nu}T^{(m)}_{\mu\nu}=\tilde g^{\mu\nu}\tilde T^{(m)}_{\mu\nu},
\label{eq:T-relations}
\end{equation}
so the trace is frame-invariant.

Because $S_m$ depends on $g_{\mu\nu}$ only through $\tilde g_{\mu\nu}=A^2(\phi)g_{\mu\nu}$, the matter sector is not separately conserved with respect to the Einstein-frame connection. Using $\nabla_\mu G^{\mu\nu}=0$ together with the field equations, one obtains the exchange law
\begin{equation}
\nabla_\mu T^{(m)\mu\nu}
=
\alpha(\phi)\,T_{(m)}\,\nabla^\nu\phi.
\label{eq:div-Tm-alpha}
\end{equation}
The sign in Eq.~\eqref{eq:div-Tm-alpha} is fixed by the scalar equation below: combining $\nabla_\mu(T^{(\phi)\mu\nu}+T^{(m)\mu\nu})=0$ with the explicit divergence $\nabla_\mu T^{(\phi)\mu\nu}=(\Box\phi-V_{,\phi})\nabla^\nu\phi$ yields Eq.~\eqref{eq:div-Tm-alpha} once $\Box\phi=V_{,\phi}-\alpha T_{(m)}$ is imposed.

Varying Eq.~\eqref{eq:action} with respect to $\phi$ gives the dilaton equation of motion
\begin{equation}
\Box\phi
=
V_{,\phi}-\alpha(\phi)\,T_{(m)},
\qquad
\Box\phi\equiv g^{\mu\nu}\nabla_\mu\nabla_\nu\phi,
\qquad
V_{,\phi}\equiv \frac{dV}{d\phi}.
\label{eq:scalar-eq}
\end{equation}
With our signature, for a perfect fluid with energy density $\rho_m$ and pressure $p_m$,
\begin{equation}
T_{(m)}=-\rho_m+3p_m,
\label{eq:trace_perfectfluid}
\end{equation}
so pressureless matter satisfies $T_{(m)}=-\rho_m$, while radiation has $T_{(m)}=0$ and does not source $\phi$ through the trace. Equations \eqref{eq:einstein-eq}, \eqref{eq:div-Tm-alpha}, and \eqref{eq:scalar-eq} provide a closed Einstein-frame description of the coupled scalar-matter system, with all observable matter dynamics minimally coupled to $\tilde g_{\mu\nu}$ via Eq.~\eqref{eq:jordan-metric}.

\section{Autonomous dynamical system}
\label{sec:autonomous}

We now specialise the Einstein-frame field equations derived in Sec.~\ref{sec:dilaton-basic} to a spatially flat FLRW background and rewrite the cosmological evolution as an autonomous dynamical system. We proceed in two steps. First, we derive a closed phase-space system valid for a generic conformal coupling $A(\phi)$ and potential $V(\phi)$, keeping the Hubble rate as a dynamical variable determined self-consistently by the cosmological solution. Second, we introduce the Damour--Polyakov (DP) least-coupling expansion and obtain a simplified system valid in the neighbourhood of $\phi=\phi_\ast$, which is the regime relevant for late-time phenomenology and precision tests of gravity.

We consider
\begin{equation}
ds^2=-dt^2+a^2(t)\,d\vec x^{\,2},
\end{equation}
where $a(t)$ is the scale factor and $H\equiv \dot a/a$ is the Hubble parameter. The dilaton is homogeneous, $\phi=\phi(t)$, and the matter sector is pressureless dust with Einstein-frame energy density $\rho_m(t)$.

\subsection{Background equations in flat FLRW}
\label{subsec:flrw_background}

From the Einstein equations \eqref{eq:einstein-eq}, the independent background equations are
\begin{eqnarray}
3M_{\rm Pl}^2 H^2
&=&
\rho_m+\rho_\phi,
\label{eq:friedmann1_full}
\\[1ex]
-2M_{\rm Pl}^2 \dot H
&=&
\rho_m+\rho_\phi+p_\phi,
\label{eq:friedmann2_full}
\end{eqnarray}
where
\begin{equation}
\rho_\phi=\frac{1}{2}\dot\phi^2+V(\phi),
\qquad
p_\phi=\frac{1}{2}\dot\phi^2-V(\phi).
\end{equation}
Using $\rho_\phi+p_\phi=\dot\phi^2$, Eq.~\eqref{eq:friedmann2_full} can be written as
\begin{equation}
-2M_{\rm Pl}^2 \dot H=\rho_m+\dot\phi^2,
\label{eq:Hdot_basic}
\end{equation}
which provides a convenient closure relation once we rewrite the system in dimensionless form.

The conformal coupling implies that matter is not covariantly conserved in the Einstein frame, as in \eqref{eq:div-Tm-alpha}. For a perfect fluid,
\begin{equation}
T^{(m)}_{\mu\nu}=(\rho_m+p_m)u_\mu u_\nu+p_m g_{\mu\nu},
\qquad
u^\mu u_\mu=-1,
\end{equation}
so the trace is $T_{(m)}=-\rho_m+3p_m$. For dust $p_m=0$ and therefore $T_{(m)}=-\rho_m$. Taking the $\nu=0$ component of \eqref{eq:div-Tm-alpha} gives
\begin{equation}
\dot\rho_m+3H\rho_m=\alpha(\phi)\,\rho_m\,\dot\phi,
\label{eq:rho_m_full}
\end{equation}
which reduces to $\rho_m\propto a^{-3}$ when $\alpha(\phi)=0$.

Finally, the homogeneous scalar equation \eqref{eq:scalar-eq} becomes
\begin{equation}
\ddot\phi+3H\dot\phi+V_{,\phi}=-\alpha(\phi)\,\rho_m,
\label{eq:phi_eom_full}
\end{equation}
where we used $T_{(m)}=-\rho_m$. Equations \eqref{eq:friedmann1_full}, \eqref{eq:Hdot_basic}, \eqref{eq:rho_m_full} and \eqref{eq:phi_eom_full} form a closed and self-consistent Einstein-frame system for $(a,\phi,\rho_m)$.

\subsection{Dimensionless variables and the full autonomous system}
\label{subsec:full_autonomous}

We use the number of $e$-folds $N\equiv \ln a$ as time variable, so that $d/dt=H\,d/dN$. For a self-consistent phase-space description we keep the Hubble scale as a dynamical variable and normalise it to a finite-epoch reference value $H_0$ (e.g.\ the present Hubble parameter). We define
\begin{eqnarray}
X &\equiv& \frac{\phi}{M_{\rm Pl}},
\\[1ex]
Y &\equiv& \frac{dX}{dN}
=
\frac{\dot\phi}{M_{\rm Pl}H},
\label{eq:Y_def_full}
\\[1ex]
\Omega_m &\equiv& \frac{\rho_m}{3M_{\rm Pl}^2 H^2},
\label{eq:Omega_m_def}
\\[1ex]
h &\equiv& \frac{H}{H_0}.
\label{eq:h_def_H0}
\end{eqnarray}
With these variables, the Friedmann constraint \eqref{eq:friedmann1_full} becomes
\begin{equation}
1=\Omega_m+\frac{Y^2}{6}+\frac{V(\phi)}{3M_{\rm Pl}^2H^2}
=
\Omega_m+\frac{Y^2}{6}+\frac{V(M_{\rm Pl}X)}{3M_{\rm Pl}^2H_0^2}\,\frac{1}{h^2}.
\label{eq:friedmann_constraint_general}
\end{equation}
This form makes explicit that, prior to any DP expansion, the potential contribution enters through the dimensionless combination $V/H^2$.

From the definition of $Y$ we have
\begin{equation}
\frac{dX}{dN}=Y.
\label{eq:dyn_X_general}
\end{equation}
Differentiating $Y=\dot\phi/(M_{\rm Pl}H)$ with respect to $N$ and using \eqref{eq:phi_eom_full} gives the general evolution equation
\begin{equation}
\frac{dY}{dN}
=
-\left(3+\frac{\dot H}{H^2}\right)Y
-\frac{V_{,\phi}}{M_{\rm Pl}H^2}
-\frac{\alpha(\phi)\rho_m}{M_{\rm Pl}H^2}.
\label{eq:dyn_Y_general}
\end{equation}
From \eqref{eq:Hdot_basic} we obtain
\begin{equation}
\frac{\dot H}{H^2}
=
-\frac{\rho_m+\dot\phi^2}{2M_{\rm Pl}^2H^2}
=
-\frac{3}{2}\Omega_m-\frac{1}{2}Y^2.
\label{eq:Hdot_over_H2}
\end{equation}
Substituting \eqref{eq:Hdot_over_H2} into \eqref{eq:dyn_Y_general} yields a closed expression for $dY/dN$ once $V(\phi)$ and $\alpha(\phi)$ are specified.

The evolution equation for $\Omega_m$ follows from \eqref{eq:rho_m_full}. Using
\begin{equation}
\frac{1}{\rho_m}\frac{d\rho_m}{dN}
=
-3+\alpha(\phi)M_{\rm Pl}Y
\end{equation}
together with $d\ln H/dN=\dot H/H^2$ from \eqref{eq:Hdot_over_H2}, we obtain
\begin{equation}
\frac{d\Omega_m}{dN}
=
\Omega_m\left[-3(1-\Omega_m)+Y^2+\alpha(\phi)M_{\rm Pl}Y\right].
\label{eq:dyn_Omega_m_general}
\end{equation}

Finally, since $h\equiv H/H_0$ and $H_0$ is constant, Eq.~\eqref{eq:Hdot_over_H2} implies
\begin{equation}
\frac{dh}{dN}
=
\frac{\dot H}{H^2}\,h
=
-\left(\frac{3}{2}\Omega_m+\frac{1}{2}Y^2\right)h.
\label{eq:dyn_h_general}
\end{equation}

Equations \eqref{eq:dyn_X_general}, \eqref{eq:dyn_Y_general}--\eqref{eq:Hdot_over_H2}, \eqref{eq:dyn_Omega_m_general} and \eqref{eq:dyn_h_general}, supplemented by the Friedmann constraint \eqref{eq:friedmann_constraint_general}, provide a self-consistent autonomous formulation of the cosmological evolution. The DP approximation will correspond to expanding the functions $\alpha(\phi)$ and $V(\phi)$ around the least-coupling point and rewriting the potential term in \eqref{eq:friedmann_constraint_general} in terms of a late-time de Sitter scale.

\subsection{Damour--Polyakov expansion and asymptotic de Sitter scale}
\label{subsec:dp_expansion}

In the Damour--Polyakov mechanism the coupling vanishes at $\phi=\phi_\ast$, so that $\alpha(\phi_\ast)=0$. Close to $\phi_\ast$ we expand
\begin{eqnarray}
A(\phi)
&\simeq&
1+\frac{1}{2}A_2(\phi-\phi_\ast)^2,
\label{eq:A_expansion_DP}
\\[1ex]
\alpha(\phi)
&\equiv&
\frac{d\ln A}{d\phi}
\simeq
A_2(\phi-\phi_\ast),
\label{eq:alpha_DP}
\\[1ex]
V(\phi)
&\simeq&
V_\ast+\frac{1}{2}m^2(\phi-\phi_\ast)^2,
\label{eq:V_quadratic_DP}
\end{eqnarray}
with constants $A_2>0$, $m^2>0$, and $V_\ast>0$. The constant term $V_\ast$ sets the asymptotic de Sitter scale
\begin{equation}
H\to H_\Lambda,
\qquad
H_\Lambda^2\equiv \frac{V_\ast}{3M_{\rm Pl}^2}.
\label{eq:HLambda_def}
\end{equation}
To avoid confusion with the $H_0$-normalised quantity $h\equiv H/H_0$, we also define
\begin{equation}
\mathcal{H}\equiv \frac{H}{H_\Lambda}.
\label{eq:calH_def}
\end{equation}

Defining the displacement $\delta\phi\equiv \phi-\phi_\ast$ and substituting \eqref{eq:alpha_DP} and \eqref{eq:V_quadratic_DP} into \eqref{eq:phi_eom_full}, the scalar equation becomes, to linear order in $\delta\phi$,
\begin{equation}
\ddot{\delta\phi}+3H\dot{\delta\phi}+\left[m^2+A_2\,\rho_m(t)\right]\delta\phi=0.
\label{eq:phi_eom_delta_full}
\end{equation}
At this order the equation is linear in $\delta\phi$. The explicit source term in \eqref{eq:rho_m_full} is proportional to $\alpha(\phi)\dot\phi$ and therefore contributes only at quadratic order in $\delta\phi$.

\subsection{DP variables and simplified autonomous system}
\label{subsec:dp_autonomous}

In the DP regime it is convenient to work with the dimensionless displacement
\begin{equation}
x\equiv \frac{\delta\phi}{M_{\rm Pl}},
\qquad
y\equiv \frac{dx}{dN}=\frac{\dot\phi}{M_{\rm Pl}H},
\label{eq:xy_defs_DP}
\end{equation}
together with $\Omega_m$ and the Hubble normalisation $h\equiv H/H_0$ introduced above. In addition, we define the constant ratio between the de Sitter scale and the reference Hubble scale,
\begin{equation}
\kappa \equiv \frac{H_\Lambda}{H_0},
\qquad\Rightarrow\qquad
\frac{H_\Lambda^2}{H^2}=\frac{\kappa^2}{h^2}.
\label{eq:kappa_def}
\end{equation}
With $V(\phi)\simeq V_\ast+\frac12 m^2M_{\rm Pl}^2x^2$, the Friedmann constraint \eqref{eq:friedmann_constraint_general} becomes
\begin{equation}
1
=
\Omega_m+\frac{y^2}{6}
+\frac{\kappa^2}{h^2}
+\frac{\nu^2}{6}\,\frac{x^2}{h^2},
\qquad
\nu\equiv \frac{m}{H_0}.
\label{eq:friedmann_constraint_DP_h}
\end{equation}

The $x$-equation remains
\begin{equation}
\frac{dx}{dN}=y.
\label{eq:dyn_x_full}
\end{equation}
Using \eqref{eq:dyn_Y_general} together with \eqref{eq:Hdot_over_H2} and the DP expansions \eqref{eq:alpha_DP}--\eqref{eq:V_quadratic_DP}, one finds
\begin{equation}
\frac{dy}{dN}
=
-\left(3-\frac{3}{2}\Omega_m-\frac{1}{2}y^2\right)y
-
\left(\frac{\nu^2}{h^2}+3A_2 M_{\rm Pl}^2\,\Omega_m\right)x.
\label{eq:dyn_y_DP_h}
\end{equation}

From \eqref{eq:dyn_Omega_m_general} and $\alpha(\phi)\simeq A_2M_{\rm Pl}x$ we obtain
\begin{equation}
\frac{d\Omega_m}{dN}
=
\Omega_m\left[-3(1-\Omega_m)+y^2+A_2 M_{\rm Pl}^2\,x\,y\right].
\label{eq:dyn_Omega_m}
\end{equation}
Finally, the Hubble equation is still given self-consistently by \eqref{eq:dyn_h_general},
\begin{equation}
\frac{dh}{dN}
=
-\left(\frac{3}{2}\Omega_m+\frac{1}{2}y^2\right)h.
\label{eq:dyn_h_general_DP}
\end{equation}

Equations \eqref{eq:dyn_x_full}, \eqref{eq:dyn_y_DP_h}, \eqref{eq:dyn_Omega_m} and \eqref{eq:dyn_h_general_DP}, together with the constraint \eqref{eq:friedmann_constraint_DP_h}, define a four-dimensional autonomous system $(x,y,\Omega_m,h)$ in the neighbourhood of the Damour--Polyakov point, while keeping the cosmological expansion self-consistent through the evolution of $h(N)$.

\subsection{Linearised dynamics near the late-time attractor}
\label{subsec:linearised_DP}

Late-time phenomenology and equivalence-principle tests probe the neighbourhood of the least-coupling configuration, where
$x\simeq 0$ and $y\simeq 0$. Over the relaxation time of the $(x,y)$ sector we treat the background quantities
$\Omega_m(N)$ and $h(N)$ as quasi-static, i.e.\ we freeze them at their values at a finite epoch $N_0$.
Linearising Eqs.~\eqref{eq:dyn_x_full} and \eqref{eq:dyn_y_DP_h} about $(x,y)=(0,0)$ then yields
\begin{eqnarray}
\frac{dx}{dN}
&=&
y,
\\[1ex]
\frac{dy}{dN}
&=&
-\,C_{\rm eff}\,y
-\,\omega_0^2\,x,
\label{eq:linear_xy_epoch_h}
\end{eqnarray}
with coefficients evaluated at $N_0$,
\begin{equation}
C_{\rm eff}\equiv 3-\frac{3}{2}\Omega_{m0},
\qquad
\omega_0^2 \equiv \frac{\nu^2}{h_0^2}+3A_2M_{\rm Pl}^2\,\Omega_{m0},
\label{eq:coeffs_epoch_h_revised}
\end{equation}
where $\Omega_{m0}\equiv \Omega_m(N_0)$ and $h_0\equiv h(N_0)=H(N_0)/H_0$ (so $h_0=1$ if $N_0$ is chosen as the
reference epoch defining $H_0$).

The corresponding Jacobian eigenvalues are
\begin{equation}
\lambda_\pm
=
\frac{-C_{\rm eff}\pm\sqrt{C_{\rm eff}^2-4\omega_0^2}}{2},
\end{equation}
which determine the local approach to the attractor: for $C_{\rm eff}^2>4\omega_0^2$ the relaxation is overdamped and
$x(N)$ decays monotonically, whereas for $C_{\rm eff}^2<4\omega_0^2$ the system exhibits damped oscillations.

On cosmological time scales the background variables evolve according to Eqs.~\eqref{eq:dyn_Omega_m} and
\eqref{eq:dyn_h_general_DP}, so the coefficients in \eqref{eq:linear_xy_epoch_h} become slowly $N$-dependent along the
trajectory. At any finite epoch one generically has $x(N_0)\neq 0$; we denote this ambient displacement by
$x_{\rm env}\equiv x(N_0)$, which sets the magnitude of the equivalence-principle effects discussed in
Sec.~\ref{sec:ep_dilaton}.

\section{Non-relativistic test particle}
\label{sec:ep_dilaton}

At the level of point-particle dynamics, matter follows geodesics of the Jordan-frame metric
$\tilde g_{\mu\nu}$, while it is often convenient to rewrite the same motion in the Einstein frame.
For a test body $A$ with constant Jordan-frame rest mass $m_A$ and worldline $x^\mu(\tau)$, the action is
\begin{equation}
S_A
=
-\,m_A\int d\tilde s
=
-\,m_A\int A(\phi)\,ds,
\label{eq:pp_action}
\end{equation}
where $d\tilde s^2=\tilde g_{\mu\nu}dx^\mu dx^\nu$ and $ds^2=g_{\mu\nu}dx^\mu dx^\nu$.
In the Einstein frame this is equivalently described as a particle with field-dependent mass
\begin{equation}
m_A(\phi)=A(\phi)\,m_A,
\label{eq:mA_phi}
\end{equation}
so that gradients of $\phi$ generate an additional (scalar-mediated) force.
Defining
\begin{equation}
\alpha(\phi)\equiv \frac{d\ln A(\phi)}{d\phi},
\end{equation}
variation of \eqref{eq:pp_action} with respect to the worldline yields the Einstein-frame equation of motion
\begin{equation}
u^\mu\nabla_\mu u^\nu
=
-\,\alpha(\phi)\,
\left(g^{\nu\lambda}+u^\nu u^\lambda\right)\nabla_\lambda\phi,
\label{eq:einstein_frame_geodesic}
\end{equation}
where $u^\mu\equiv dx^\mu/ds$ and $\nabla_\mu$ is compatible with $g_{\mu\nu}$.

\medskip

In the weak-field, non-relativistic regime we take
\begin{equation}
ds^2 \simeq -(1+2\Phi_N)\,dt^2+(1-2\Phi_N)\,d\vec x^{\,2},
\end{equation}
and, for $|\dot{\mathbf r}|\ll 1$, Eq.~\eqref{eq:einstein_frame_geodesic} reduces to
\begin{equation}
\ddot{\mathbf r}
=
-\,\nabla\Phi_N
-\,\alpha(\phi)\,\nabla\phi,
\label{eq:acc_einstein_NR}
\end{equation}
where dots denote derivatives with respect to $t$.
The first term is the standard Newtonian acceleration, while the second term is the fifth-force contribution.

Multiplying \eqref{eq:acc_einstein_NR} by the inertial mass $m_{\rm in}$ (defined operationally as the coefficient of
$\ddot{\mathbf r}$) gives
\begin{equation}
m_{\rm in}\,\ddot{\mathbf r}
=
-\,m_{\rm in}\,\nabla\Phi_N
-\,m_{\rm in}\,\alpha(\phi)\,\nabla\phi.
\label{eq:newton2_dilaton}
\end{equation}
The extra scalar term implies that, in the Einstein-frame description, the motion is not sourced solely by $\Phi_N$
when $\nabla\phi\neq 0$.

\medskip

A passive gravitational mass can be introduced operationally by rewriting the total force in Newtonian form,
\begin{equation}
m_{\rm in}\,\ddot{\mathbf r}
\equiv
-\,m_g^{(p)}\,\nabla\Phi_N.
\label{eq:def_mg_passive}
\end{equation}
Comparing \eqref{eq:newton2_dilaton} and \eqref{eq:def_mg_passive} yields
\begin{equation}
\frac{m_g^{(p)}}{m_{\rm in}}
=
1
+
\frac{\alpha(\phi)\,\nabla\phi\cdot\hat{\mathbf r}}{\nabla\Phi_N\cdot\hat{\mathbf r}},
\label{eq:mg_over_min_general}
\end{equation}
where $\hat{\mathbf r}$ denotes the radial direction.
This ratio should be understood as a compact parametrisation of the fifth-force contribution: it does not imply a
fundamental change of masses, but rather encodes that the same Newtonian potential produces a different acceleration
once the scalar force is present.

\medskip

To connect \eqref{eq:mg_over_min_general} with the phase-space variables of Sec.~\ref{sec:autonomous}, we relate the locally
sourced scalar profile to the Newtonian potential.
Consider a weak-field configuration sourced by a non-relativistic density $\rho(\mathbf r)$.
Write $\phi=\phi_{\rm env}+\varphi$, where $\phi_{\rm env}$ is the ambient value of the field (set by the cosmological background
in the environment of interest) and $\varphi$ is the local perturbation generated by the source.
Assuming that the scalar is effectively long-range on the relevant scales (so that an effective mass term can be neglected)
and linearising the scalar equation around $\phi_{\rm env}$ gives
\begin{equation}
\nabla^2\varphi
\simeq
\alpha_{\rm env}\,\rho,
\qquad
\alpha_{\rm env}\equiv \alpha(\phi_{\rm env}),
\label{eq:poisson_phi_linear}
\end{equation}
where for dust we used $T_{(m)}\simeq -\rho$ in \eqref{eq:scalar-eq} and kept $\alpha(\phi)$ fixed at $\alpha_{\rm env}$ to this
order.
The Newtonian potential satisfies the Poisson equation
\begin{equation}
\nabla^2 \Phi_N
=
\frac{\rho}{2M_{\rm Pl}^2}.
\label{eq:poisson_phiN}
\end{equation}
Eliminating $\rho$ between \eqref{eq:poisson_phi_linear} and \eqref{eq:poisson_phiN} yields
\begin{equation}
\nabla^2\varphi
=
2\,\alpha_{\rm env}\,M_{\rm Pl}^2\,\nabla^2\Phi_N.
\end{equation}
Imposing the same boundary conditions for $\varphi$ and $\Phi_N$ at spatial infinity implies
\begin{equation}
\nabla\phi
=
\nabla\varphi
\simeq
2\,\alpha_{\rm env}\,M_{\rm Pl}^2\,\nabla\Phi_N,
\label{eq:gradphi_gradPhi}
\end{equation}
up to corrections associated with finite-range effects (e.g.\ Yukawa suppression) or screening.

Substituting \eqref{eq:gradphi_gradPhi} into \eqref{eq:mg_over_min_general} and evaluating $\alpha(\phi)$ at $\phi_{\rm env}$
gives
\begin{equation}
\frac{m_g^{(p)}}{m_{\rm in}}
\simeq
1+2\,\bigl(M_{\rm Pl}\alpha_{\rm env}\bigr)^2,
\label{eq:mg_over_min_alpha2}
\end{equation}
showing that the magnitude of the fifth-force contribution is governed by the ambient coupling.

\medskip

The quantity that is directly constrained by universality-of-free-fall experiments is the E\"otv\"os parameter,
defined for two test bodies $A$ and $B$ falling in the gravitational field of a source $S$ as \cite{Will2014,CassiniBertotti2003}:
\begin{equation}
\eta_{AB}\equiv \frac{2|a_A-a_B|}{a_A+a_B},
\label{eq:eotvos_def}
\end{equation}
where $a_A$ and $a_B$ are the measured magnitudes of the accelerations.
In scalar-tensor theories, the acceleration contains both the Newtonian part and a scalar contribution. If the scalar coupling
is universal, in the sense that both bodies share the same effective coupling to $\phi$ in the given environment, then the
scalar force rescales the common free-fall acceleration but does not produce differential acceleration, and one finds
$\eta_{AB}=0$ at leading order.
A nonzero $\eta_{AB}$ arises when the effective couplings are composition dependent, i.e.\ when different bodies acquire
different scalar charges (for example through dilaton-type couplings to the Standard Model sector)\cite{Microscope2017, Microscope2022, CliftonReview}. In that case, one may
parametrise the non-relativistic acceleration of $A$ in the field of $S$ as
\begin{equation}
\mathbf a_A
=
-\,\nabla\Phi_N\left[1+2M_{\rm Pl}^2\,\alpha_A(\phi_{\rm env})\,\alpha_S(\phi_{\rm env})\right],
\end{equation}
so that, for small couplings, the E\"otv\"os parameter is approximately
\begin{equation}
\eta_{AB}\simeq 2M_{\rm Pl}^2\,\alpha_S(\phi_{\rm env})\left[\alpha_A(\phi_{\rm env})-\alpha_B(\phi_{\rm env})\right],
\label{eq:eotvos_smallcoupling}
\end{equation}
with all couplings evaluated at the ambient value relevant for the experiment.
This makes explicit the role of the cosmological attractor: the autonomous system determines the evolution of $\phi(N)$ and
therefore the ambient couplings $\alpha(\phi_{\rm env})$ that enter laboratory and Solar System tests.

\medskip

Near the Damour-Polyakov point, the coupling is linear in the displacement from $\phi_*$,
\begin{equation}
\alpha(\phi)\simeq A_2(\phi-\phi_*),
\end{equation}
therefore
\begin{equation}
\alpha_{\rm env}
\simeq
A_2(\phi_{\rm env}-\phi_*)
=
A_2 M_{\rm Pl}\,x_{\rm env},
\label{eq:alpha_env_x}
\end{equation}
where $x_{\rm env}\equiv(\phi_{\rm env}-\phi_*)/M_{\rm Pl}$ is the phase-space displacement of the cosmological trajectory.
Combining \eqref{eq:mg_over_min_alpha2} and \eqref{eq:alpha_env_x} yields
\begin{equation}
\frac{m_g^{(p)}}{m_{\rm in}}
\simeq
1
+
2A_2^2 M_{\rm Pl}^4\,x_{\rm env}^2.
\label{eq:mg_over_min_x}
\end{equation}
This result makes the bridge to Sec.~\ref{sec:autonomous} explicit: the dynamical system determines $x(N)$, and the ambient
value $x_{\rm env}$ relevant for a given experiment is obtained by evaluating the cosmological solution at the epoch and
environment of interest. Since the fixed point $x\to 0$ is approached only asymptotically, one generally has $x_{\rm env}\neq 0$
at finite time, and Eq.~\eqref{eq:mg_over_min_x} shows how even a small residual displacement produces a nonzero fifth-force
contribution.
Moreover, the linearised subsystem in Sec.~\ref{sec:autonomous} fixes the local relaxation of $x(N)$ near $(0,0)$ through the Jacobian eigenvalues $\lambda_\pm$. As a result, the rate at which $x_{\rm env}$ is driven toward zero---and hence the size of the predicted effects in Eqs.~\eqref{eq:mg_over_min_x} and \eqref{eq:eotvos_smallcoupling} at any finite epoch---is determined by the stability properties of the attractor together with the trajectory's initial conditions.

This differs from chameleon scenarios, where the suppression of scalar-mediated forces is primarily environmental: the effective potential depends explicitly on the ambient density, which typically makes the field heavy in high-density regions and leads to a reduced range and/or a thin-shell behaviour around extended bodies. In that case, the effective coupling probed by laboratory or Solar-System tests is controlled by local properties of the environment and of the source—density profiles, gravitational potential depth, and screening radii—so that bounds are naturally expressed in terms of screening conditions rather than a cosmological offset variable \cite{KhouryWeltman2004_PRD,KhouryWeltman2004_PRL,Brax2004_ChameleonDE,Brax2004_ChameleonDE_JCAP,Gannouji2010_Chameleon,BraxMotaShaw2011,BraxBurrage2014_review}.

Symmetron-type models provide a complementary contrast: here screening is associated with symmetry restoration in dense environments, which drives the field toward a symmetry-preserving configuration and suppresses the matter coupling in screened regions, while allowing a nonzero field expectation value and stronger coupling in low-density environments. Disformal screening further illustrates that the relevant suppression can be controlled by derivative interactions and matter-dependent effective metrics, again tying the phenomenology to local environmental data. In this sense, chameleon/symmetron/disformal mechanisms implement screening through environment-driven modifications of the local scalar profile, whereas in the DP regime considered here the dominant control parameter is $x_{\rm env}$—a cosmological residual that is predicted by the autonomous dynamics and whose decay rate is governed by the stability data of the late-time attractor \cite{HinterbichlerKhoury2010_PRL,Hinterbichler2011_SymmetronCosmology,Davis2012_StructureSymmetron,Koivisto2012_ScreeningDisformal,Hagala2016_DisformalSymmetron,BraxSymmetronReview2015}.

\section{Phase portrait and discussion of some results}
\label{sec:numerics}

We consider the \emph{linearised} Damour--Polyakov (DP) dynamics in the $(x,y)$ plane, obtained by expanding the
full system about the least--coupling configuration $(x,y)=(0,0)$ at a reference epoch $N_0$.
This approximation isolates the local relaxation mechanism that is most relevant for phenomenology: in the DP
regime the ambient scalar--matter coupling scales as $\alpha_{\rm env}\propto x_{\rm env}$, so the decay of $x$
directly controls the suppression of fifth forces and deviations from the equivalence principle discussed in
Sec.~\ref{sec:ep_dilaton}.
\footnote{A dedicated numerical investigation of the full DP cosmological system $(x,y,\Omega_m,h)$ is being
developed separately, focusing on the global phenomenology, including parameter regions compatible with late--time
acceleration and possible connections with cosmological tensions.}

\begin{figure}[t]
\centering
\includegraphics[width=0.450\textwidth]{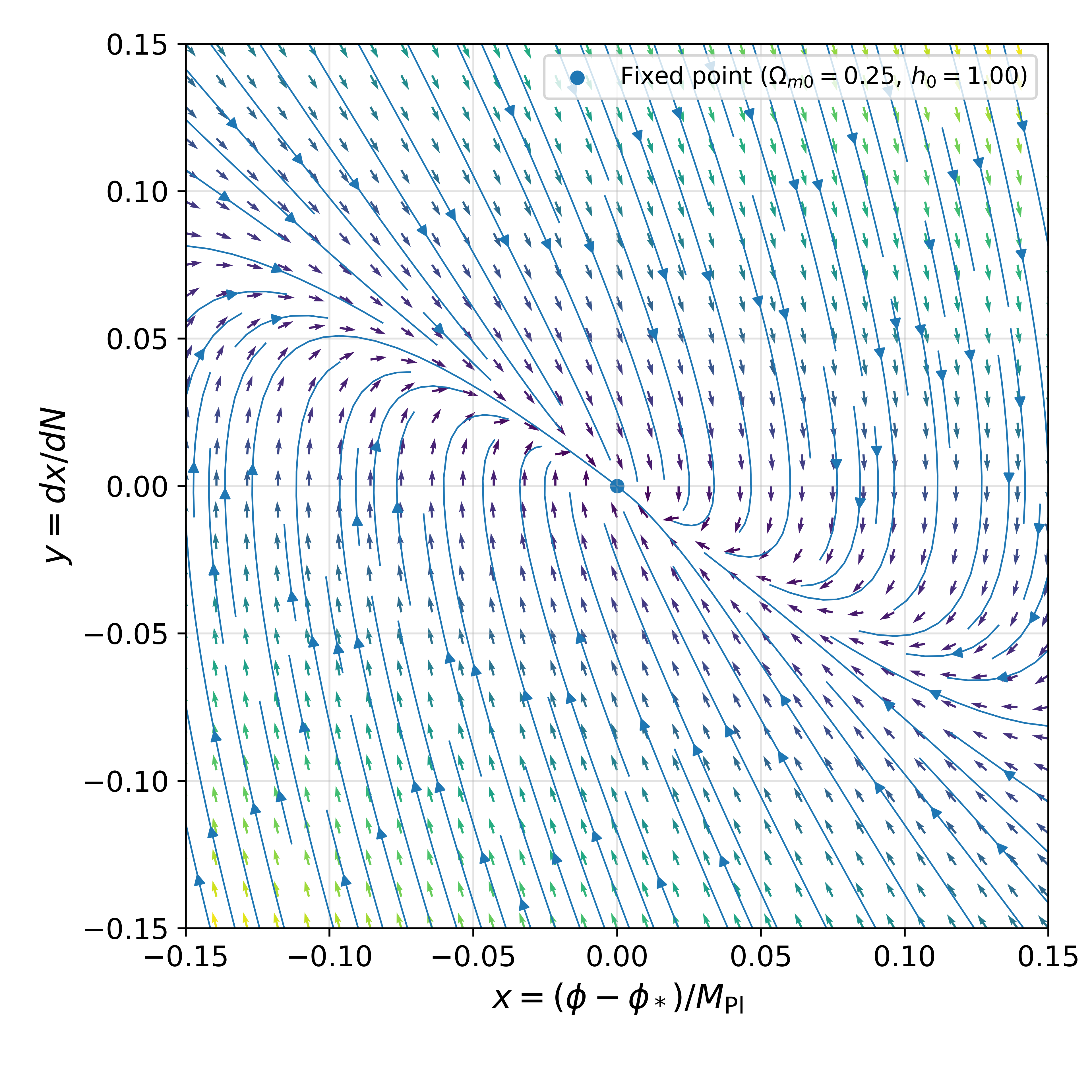}
\includegraphics[width=0.450\textwidth]{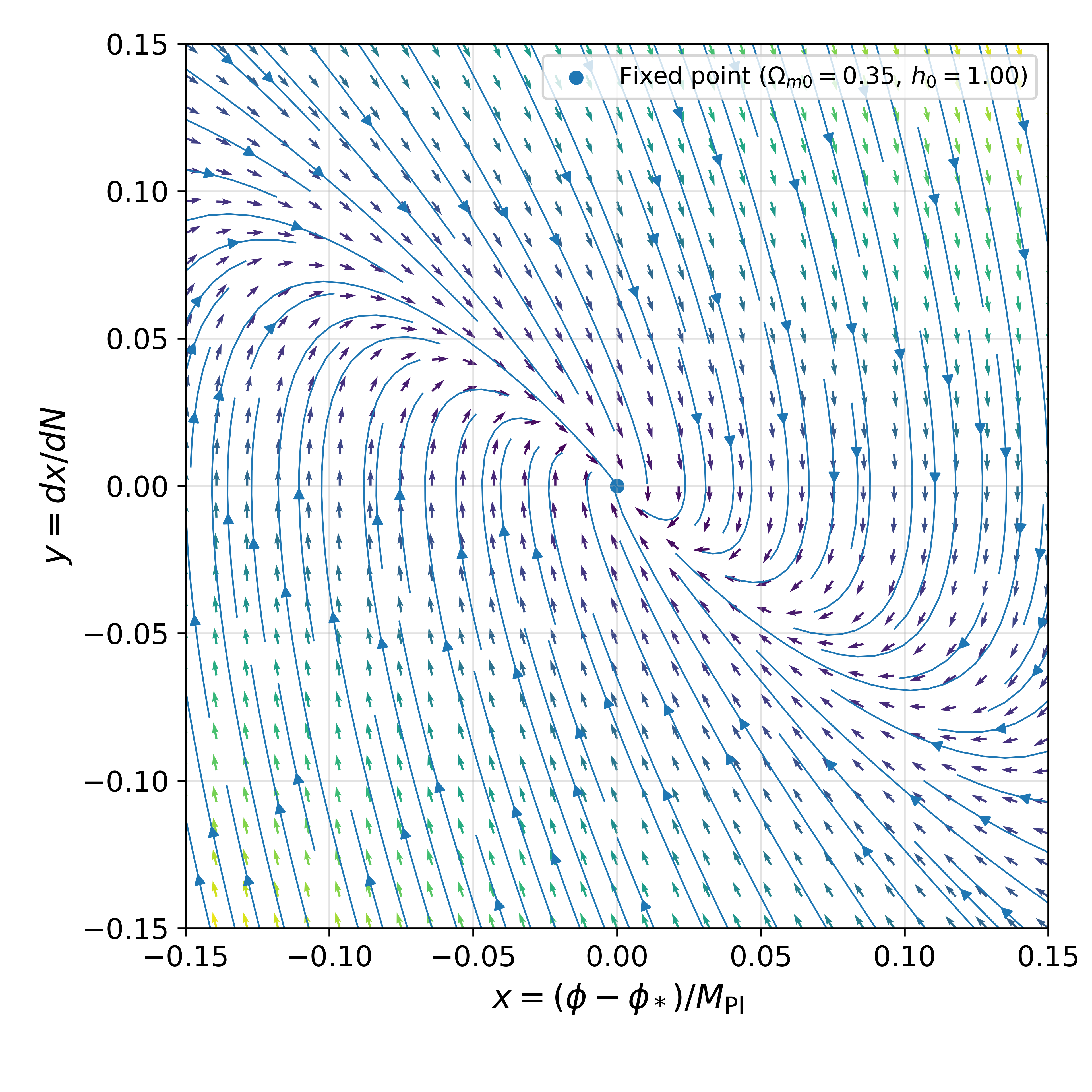}
\caption{
Quasi-static phase portraits of the linearised Damour--Polyakov subsystem in the $(x,y)$ plane for two background matter fractions: (a) $\Omega_{m0}=0.25$ and (b) $\Omega_{m0}=0.32$, both with $h_0=1.0$. The fixed point at $(0,0)$ corresponds to the least-coupling configuration.
}
\label{fig1}
\end{figure}

Figure~\ref{fig1} shows the corresponding phase portraits for two representative matter fractions,
$\Omega_{m0}=0.35$ (right) and $\Omega_{m0}=0.25$ (left), both evaluated at $h_0=1$.
In each case the origin is an attractor and trajectories approach it through damped spirals, demonstrating that
the least--coupling state is locally stable over this range of $\Omega_{m0}$.
Varying $\Omega_{m0}$ mainly shifts the relative weight of the effective friction and restoring terms, changing
the damping time and oscillation scale while preserving the qualitative topology of the flow.

\begin{figure}[t]
\centering
\includegraphics[width=0.700\textwidth]{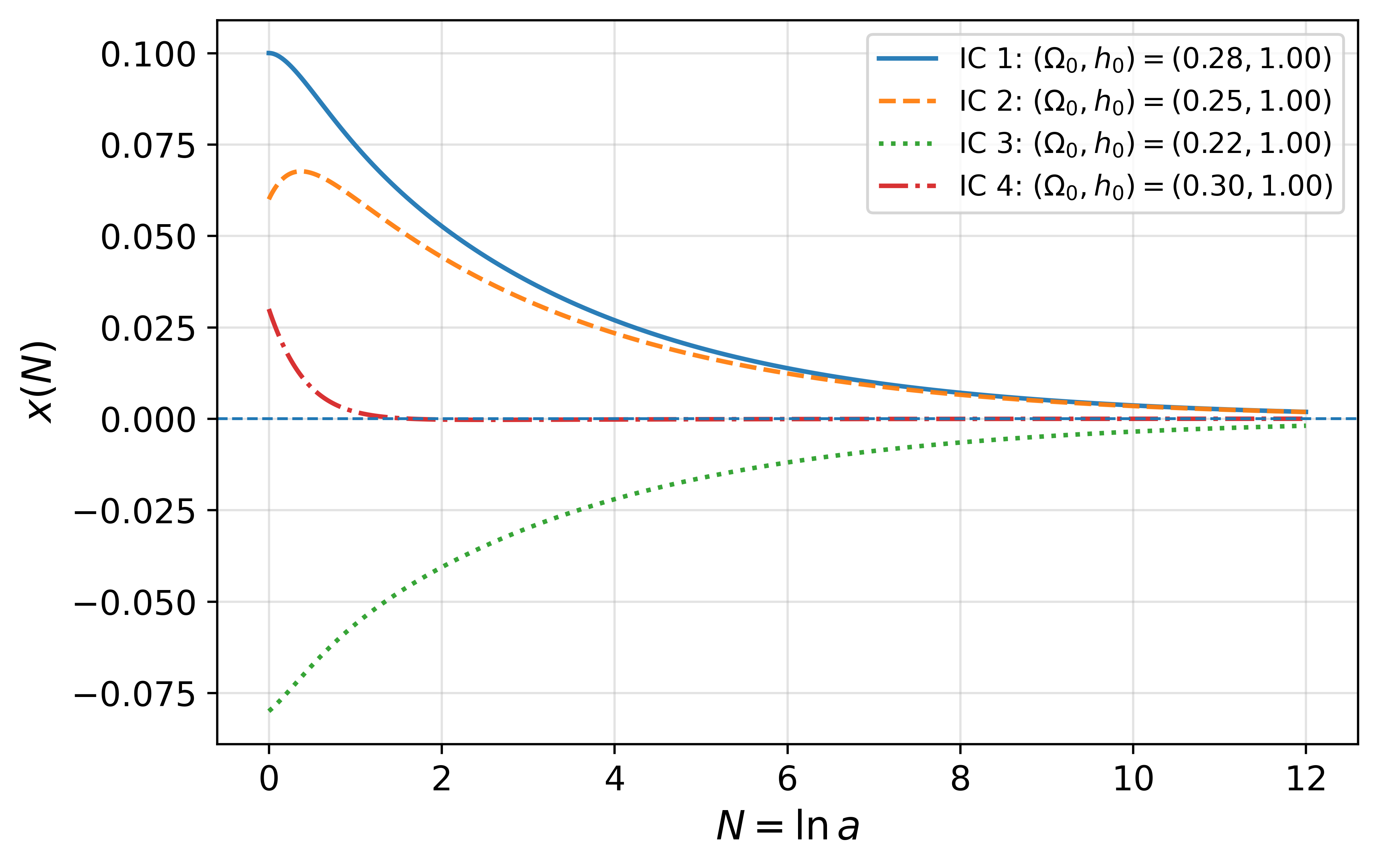}
\caption{
Evolution of the dimensionless scalar displacement $x(N)=(\phi-\phi_\ast)/M_{\rm Pl}$ as a function of $N=\ln a$
for representative initial conditions $(\Omega_0,h_0)$.
In all cases $x$ is damped and relaxes towards $x\to 0$, approaching the least--coupling configuration.
Since the ambient matter coupling in the Damour--Polyakov regime scales as $\alpha_{\rm env}\propto x$,
the decay of $x(N)$ implies a progressive suppression of composition-dependent fifth forces and
deviations from the equivalence principle at late times.
}

\label{fig2}
\end{figure}

The time-domain behaviour in Fig.~\ref{fig2} provides the complementary perspective: for several initial conditions
$(\Omega_0,h_0)$, the scalar displacement $x(N)=(\phi-\phi_\ast)/M_{\rm Pl}$ relaxes towards $x\to 0$ as $N$ grows.
Since $\alpha_{\rm env}\propto x_{\rm env}$, this relaxation implies a progressive reduction of composition-dependent
interactions and deviations from the equivalence principle at late times, with only weak dependence on moderate
changes in the initial matter fraction.

Near $(x,y)=(0,0)$ the dynamics is controlled by the Jacobian eigenvalues $\lambda_\pm$, which depend on the effective
friction coefficient $C_{\rm eff}$ and effective frequency $\omega_0$ defined in Eq.~\eqref{eq:coeffs_epoch_h_revised}.
Real $\lambda_\pm$ correspond to overdamped decay, whereas complex $\lambda_\pm$ yield damped oscillations; in both
cases the phase portrait provides a direct geometric measure of how efficiently $x_{\rm env}$ is driven to small
values and, consequently, how rapidly deviations from the equivalence principle are suppressed.

\section{Conclusions}

In this work we developed a dynamical-systems framework to analyse a string--inspired dilaton cosmology in the
Damour-Polyakov (DP) regime and to make explicit how the cosmological evolution controls deviations from the
equivalence principle. Expanding the conformal factor and the scalar potential around the least-coupling point
$\phi=\phi_\ast$, we obtained a closed and self--consistent autonomous system in terms of $(x,y,\Omega_m,h)$, where
$x\equiv (\phi-\phi_\ast)/M_{\rm Pl}$ and $y\equiv dx/dN$. The resulting phase space admits a stable fixed point at
$(x,y)=(0,0)$, approached only asymptotically by cosmological trajectories.

The main contribution of the paper is to provide a transparent dynamical link between the background cosmology and
equivalence--principle physics: at any finite epoch the cosmological solution retains a residual displacement
$x_{\rm env}\neq 0$, which sets the ambient coupling
$\alpha_{\rm env}\simeq A_2 M_{\rm Pl}\,x_{\rm env}$ and therefore fixes the strength of the scalar force in the
nonrelativistic limit. In particular, the fifth-force contribution can be encoded as an effective rescaling of the
Newtonian response, leading to the characteristic scaling
$(m_g^{(p)}/m_{\rm in}-1)\propto \alpha_{\rm env}^2$. Thus, the relevance of the DP mechanism is not merely the
existence of an attractor, but the finite--time distance to it along the cosmological history, which directly
determines the level of deviations from the equivalence principle.

To quantify how efficiently the cosmological evolution suppresses $x_{\rm env}$, we linearised the $(x,y)$ sector at
a finite reference epoch and expressed the local relaxation in terms of the Jacobian eigenvalues $\lambda_\pm$.
These eigenvalues control both the damping rate and the overdamped or oscillatory character of the approach, while
stability is guaranteed by ${\rm Re}(\lambda_\pm)<0$. The resulting picture is illustrated by the phase portraits
and time--domain solutions presented in Sec.~\ref{sec:numerics}.

It is important to stress the scope of the present paper. Our goal here was to establish the dynamical mechanism
linking cosmological relaxation to deviations from the equivalence principle, rather than to derive observational
constraints. A natural next step is a dedicated numerical exploration of the full four--dimensional system
$(x,y,\Omega_m,h)$, mapping viable regions of parameter space and extracting realistic histories
$\alpha_{\rm env}(N)$ to be confronted with laboratory and Solar--System tests, including composition--dependent
scalar charges and finite--range effects. Such an extension would connect the local linearised description
developed here to the global cosmological phenomenology of the complete DP model \cite{MarchBertolamiMuccino2024,Tino2020EPReview,Mancini2025EquivGravities,Onofrio2025WEPTests}.

\section*{Acknowledgments}

I would like to thank Maria Margarita and Miguel Amado for their continuous inspiration and support, and the Foundation for Research Support of Espírito Santo (FAPES) for the partial support for the present work.

\end{document}